# Semi-Blind Source Separation for Nonlinear Acoustic Echo Cancellation

Guoliang Cheng, Lele Liao, Hongsheng Chen, and Jing Lu

*Abstract*—The mismatch between the numerical and actual nonlinear models is a challenge to nonlinear acoustic echo cancellation (NAEC) when the nonlinear adaptive filter is utilized. To alleviate this problem, we combine a basis-generic expansion of the memoryless nonlinearity into semi-blind source separation (SBSS). By regarding all the basis functions of the far-end input signal as the known equivalent reference signals, an SBSS updating algorithm is derived following the constrained scaled natural gradient strategy. Unlike the commonly utilized adaptive algorithm, the proposed SBSS is based on the independence between the near-end signal and the reference signals, and is less sensitive to the mismatch of nonlinearity between the numerical and actual models. Experimental results show that the proposed method outperforms conventional methods in terms of echo return loss enhancement (ERLE) and near-end speech quality evaluated by perceptual evaluation of speech quality (PESQ) and short-time objective intelligibility (STOI).

## I. INTRODUCTION

LINEAR acoustic echo cancellation (AEC) assumes that the far-end input signal is linearly convolved with the echo path to give the echo signal, and usually the linear adaptive filtering method can be used to estimate the echo path and eliminate the echo signal [1]–[3]. Semi-blind source separation (SBSS), rooted from blind source separation (BSS) [4], can also be used for linear AEC [5]–[7]. The SBSS method was first proposed in [8] and was successfully implemented in [5] as a combination of a multichannel BSS and a single-channel AEC in the frequency domain. It was subsequently shown in [6] and [7] that BSS and multichannel AEC can be combined effectively, resulting in an SBSS without double-talk detection. SBSS were also proven to be able to estimate the echo path during double-talk [9]–[11].

In practical applications, nonlinearity is always inevitable especially in the portable devices such as smartphones and laptops which use miniature loudspeakers. Therefore, the

This work was supported by the National Natural Science Foundation No. 11874219 of China.

Guoliang Cheng, Lele Liao, Hongsheng Chen, and Jing Lu are with Key Laboratory of Modern Acoustics, Institute of Acoustics, Nanjing University, Nanjing 210093, China, NJU-Horizon Intelligent Audio Lab, Horizon Robotics, Beijing 100094, China, and Nanjing Institute of Advanced Artificial Intelligence, Nanjing 210014, China (e-mail: chengguoliang@smail.nju.edu.cn; liaolele@smail.nju.edu.cn; hschen@smail.nju.edu.cn; lujing@nju.edu.cn).

nonlinear acoustic echo cancellation (NAEC) is preferred for better echo cancellation performance. Over the years, a large number of models have been employed to describe the nonlinearity in the NAEC system, such as Wiener-Hammerstein model [12], Volterra model [13], polynomial saturation [14], and neural networks [15]. The memoryless nonlinearity has been shown to model well the nonlinear distortion of loudspeakers [16], in which the echo path can be decomposed into a cascade structure of a nonlinear model and linear echo path [17], [18]. The combination of adaptive filters, in which linear and nonlinear elements are adapted separately using two different adaptive filters in parallel, was proposed in [19] and further improved by [20], [21]. In contrast, the linear echo path and the nonlinearity can be estimated by the well-known Kalman filter, which shows a fast adaptation with a better performance [22]–[24] than both the least mean square (LMS) and the recursive least squares (RLS) algorithms [1]. A multichannel state-space frequency-domain adaptive filter (MCSSFDAF) was devised in [23] to approach the NAEC problem by considering a basis-generic expansion of the memoryless nonlinearity. The method proposed in [23] was further extended and implemented in microphone array [25]. Moreover, a dual-stage multichannel Kalman (DualStage-MCK) filter for NAEC was recently proposed in [26] by using two serially cascaded adaptive filters.

The performance of NAEC largely depends on the accuracy of the numerical nonlinear model since the adaptive filtering used in AEC in essence aims at identifying the transfer function between the excitation of the loudspeaker and the captured signal of the microphone. In practical applications, the mismatch between the numerical and actual nonlinear models is inevitable and this will be detrimental to the NAEC system and lead to deteriorated performance of those based on adaptive filtering.

We note that SBSS is usually designed based on the assumption of independence between the reference signal and the near-end signal [5]–[11], and is theoretically less sensitive to the mismatch between the numerical and actual transfer functions. However, current SBSS is used to solve the linear AEC problem, and cannot be applied in the NAEC system. In this letter, we modify the SBSS to make it suitable to be implemented in NAEC. The basis-generic expansion of the memoryless nonlinearity is applied to the reference signal first and the expansion coefficients are merged into the echo path. By regarding each basis function of the far-end input signal as the known equivalent reference signal, an updating process is designed following the constrained scaled natural gradient strategy [7], [27] commonly used in linear SBSS. By combining the merits of the SBSS and the nonlinear expansion, the proposed method can have better performance in practical applications where the nonlinear model deviates from the actual model.

## II. SBSS Model for NAEC

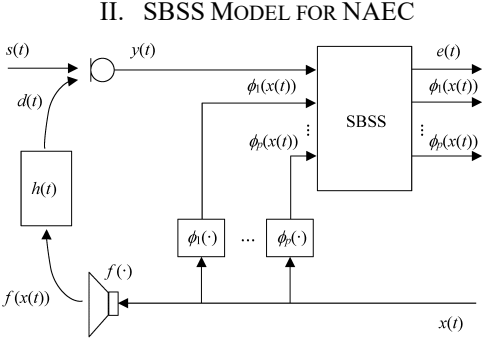

Fig. 1. SBSS model for NAEC in the presence of the memoryless nonlinearity.

The proposed SBSS model for the NAEC system is depicted in Fig. 1. The loudspeaker nonlinearity is modeled as a memoryless nonlinear function $f(\cdot)$, which transforms the far-end input signal $x(t)$ with time index $t$ into the nonlinearly mapped input signal $f(x(t))$. The signal $f(x(t))$ gets linearly convolved with the echo path $h(t)$ to obtain the echo signal $d(t)$. The near-end signal $s(t)$ is then superimposed on the echo signal $d(t)$, resulting in the microphone signal $y(t)$ as

$$y(t) = d(t) + s(t) = h(t) * f(x(t)) + s(t). \quad (1)$$

In practical applications, the actual nonlinear model is unknown, and a basis-generic expansion of the nonlinearly mapped input signal $f(x(t))$ is often utilized as [22], [23]

$$f(x(t)) = \sum_{i=1}^{p} a_i \phi_i(x(t)), \quad (2)$$

where $\phi_i(\cdot)$ is the $i$th-order basis function, $a_i$ is the corresponding coefficient, and $p$ is the expansion order. The odd power series [26] described as

$$\phi_i(x(t)) = x^{2i-1}(t), \quad i = 1, 2, \ldots, p, \quad (3)$$

is preferred because it can achieve the best performance when compared to other expansions, such as Legendre or Fourier expansions and the even power series, with the same order. Substituting (2) into (1) yields

$$y(t) = h(t) * \left[\sum_{i=1}^{p} a_i \phi_i(x(t))\right] + s(t). \quad (4)$$

By merging the expansion coefficients $a_i$ into the echo path $h(t)$, (4) can be expressed as

$$y(t) = \sum_{i=1}^{p} h'_i(t) * \phi_i(x(t)) + s(t), \quad (5)$$

where $h'_i(t)$ represents the echo path corresponding to the $i$th-order basis function as

$$h'_i(t) = a_i h(t). \quad (6)$$

By using short-time Fourier transform (STFT), the frequency-domain representation of (5) can be obtained as

$$Y(k,n) = \sum_{i=1}^{p} H_i(k,n) X_{\phi,i}(k,n) + S(k,n), \quad (7)$$

where $Y(k, n)$, $H_i(k, n)$, $X_{\phi,i}(k, n)$, and $S(k, n)$ are the frequency-domain representations of $y(t)$, $h'_i(t)$, $\phi_i(x(t))$, and $s(t)$ respectively with the frequency index $k$ and the frame index $n$. Combine $X_{\phi,i}(k, n)$ with $Y(k, n)$ and $S(k, n)$ respectively into the vector forms as

$$\mathbf{y}(k,n) = [Y(k,n), X_{\phi,1}(k,n), \ldots, X_{\phi,p}(k,n)]^T, \quad (8)$$

$$\mathbf{s}(k,n) = [S(k,n), X_{\phi,1}(k,n), \ldots, X_{\phi,p}(k,n)]^T, \quad (9)$$

then the matrix form of (7) can be represented as

$$\mathbf{y}(k,n) = \mathbf{H}(k,n)\mathbf{s}(k,n), \quad (10)$$

where $\mathbf{H}(k, n)$ is a mixing matrix of size $(p + 1) \times (p + 1)$ in block formulation as

$$\mathbf{H}(k,n) = \begin{bmatrix} 1 & \mathbf{h}^T(k,n) \\ \mathbf{0}_{p \times 1} & \mathbf{I}_p \end{bmatrix}, \quad (11)$$

with $\mathbf{0}_{p \times 1}$ a zero vector of size $p \times 1$, $\mathbf{I}_p$ an identity matrix of size $p \times p$, and $\mathbf{h}(k, n)$ a mixing vector of size $p \times 1$ as

$$\mathbf{h}(k,n) = [H_1(k,n), \ldots, H_p(k,n)]^T. \quad (12)$$

Since $x(t)$ is the known input signal, $\phi_i(x(t))$ and $X_{\phi,i}(k, n)$ are also known. Regarding $X_{\phi,i}(k, n)$ as the reference signal, the unknown near-end signal $S(k, n)$ can be extracted using the SBSS method. The demixing process is described as

$$\mathbf{e}(k,n) = \mathbf{W}(k,n)\mathbf{y}(k,n), \quad (13)$$

where $\mathbf{e}(k, n)$ is the estimated vector of size $(p + 1) \times 1$ and $\mathbf{W}(k, n)$ is the demixing matrix of size $(p + 1) \times (p + 1)$. They have the forms as

$$\mathbf{e}(k,n) = [E(k,n), X_{\phi,1}(k,n), \ldots, X_{\phi,p}(k,n)]^T, \quad (14)$$

$$\mathbf{W}(k,n) = \begin{bmatrix} 1 & \mathbf{w}^T(k,n) \\ \mathbf{0}_{p \times 1} & \mathbf{I}_p \end{bmatrix}, \quad (15)$$

where $E(k, n)$ is the estimate of the near-end signal $S(k, n)$ and $\mathbf{w}(k, n)$ is the demixing vector of size $p \times 1$.

One may argue that all the $X_{\phi,i}(k, n)$ are transformed from the same reference signal $x(t)$; therefore they do not satisfy the independence assumption, which is the basis of the BSS method. However, for the SBSS used in NAEC, the near-end signal is independent of the reference signals, and the mixing and demixing matrices are both constrained, so that the near-end signal can still be effectively recovered. The feasibility of SBSS with linearly dependent reference signals has already been verified in multichannel linear AEC [7].

## III. Online SBSS Algorithm

The demixing matrix in (13) can be optimized with an online SBSS algorithm based on the real-time natural gradient independent vector analysis (IVA) [28]–[31]. Following similar derivation in online IVA [29], the update rule of $\mathbf{W}(k, n)$ is given as

$$\mathbf{W}(k,n+1) = \mathbf{W}(k,n) + \eta\left[\mathbf{I}_{p+1} - \Psi(\mathbf{e}(k,n))\mathbf{e}^H(k,n)\right]\mathbf{W}(k,n), \quad (16)$$

where $\eta$ is learning rate, $(\cdot)^H$ denotes Hermitian transpose, and the nonlinear function $\Psi(\cdot)$ is known as a multivariate score function. A typical form of this multivariate score function is from the dependent multivariate super-Gaussian distribution in [28] as

$$\Psi(\mathbf{e}(k,n)) = [\Psi(e_1(k,n)),\ldots,\Psi(e_{p+1}(k,n))]^T, \quad (17)$$

with

$$\Psi(e_j(k,n)) = \frac{e_j(k,n)}{\sqrt{\sum_{k=1}^{K}|e_j(k,n)|^2}}, \quad (18)$$

where $e_j(k, n)$ represents the $j$th element of the vector $\mathbf{e}(k, n)$ and $K$ is the number of frequency bins. To obtain a stable algorithm while preserving the constrained structure of the demixing matrix in (15), we further use the constrained scaled natural gradient strategy [7], [27] and the update equations are expressed as

$$\Delta\mathbf{W}(k,n) = \left[\mathbf{I}_{p+1} - \frac{1}{d(k,n)}\Psi(\mathbf{e}(k,n))\mathbf{e}^H(k,n)\right]\mathbf{W}(k,n), \quad (19)$$

$$\Delta\mathbf{W}_{2:p+1,:}(k,n) = \mathbf{O}_{p\times(p+1)}, \quad (20)$$

$$\mathbf{W}(k,n+1) = c(k,n)[\mathbf{W}(k,n) + \eta\Delta\mathbf{W}(k,n)], \quad (21)$$

$$\mathbf{W}_{1,:}(k,n+1) = \frac{\mathbf{W}_{1,:}(k,n+1)}{\mathbf{W}_{1,1}(k,n+1)}, \quad (22)$$

$$\mathbf{W}_{2:p+1,2:p+1}(k,n+1) = \mathbf{I}_p, \quad (23)$$

where $\Delta\mathbf{W}(k, n)$ is the updating term of $\mathbf{W}(k, n)$, $\mathbf{O}_{p\times(p+1)}$ represents a zero matrix of size $p \times (p + 1)$, $\Delta\mathbf{W}_{2:p+1,:}(k, n)$ represents a matrix composed of the 2nd to $(p + 1)$-th rows of the matrix $\Delta\mathbf{W}(k, n)$, $\mathbf{W}_{1,:}(k, n + 1)$ represents the first row of the matrix $\mathbf{W}(k, n + 1)$, $\mathbf{W}_{1,1}(k, n + 1)$ represents the element in the first row and first column of the matrix $\mathbf{W}(k, n + 1)$, $\mathbf{W}_{2:p+1,2:p+1}(k, n + 1)$ represents a square matrix of size $p \times p$ at the bottom right corner of the matrix $\mathbf{W}(k, n + 1)$, and $d(k, n)$ and $c(k, n)$ are the scaling factors, which are computed as in [27].

The proposed SBSS algorithm is based on the independence between the near-end signal and the reference signals. Thus it is less sensitive to the mismatch of nonlinearity between the numerical and actual models.

## IV. SIMULATIONS AND EXPERIMENTS

In order to verify the effectiveness of the proposed algorithm, we compare the performance of the SBSS algorithm with that of the state-of-the-art NAEC algorithms based on submatrix-diagonal MCSSFDAF (SD-MCSSFDAF) [23] and DualStage-MCK [26] using both simulated and real captured data. Exemplary audio samples are available online at https://github.com/ChengGuoliang0/audio-samples.

### A. Simulations

We consider two types of nonlinear mappings to model the memoryless loudspeaker nonlinearity: hard clipping and soft saturation [23], [25]. The hard clipping model is expressed as

$$f(x(t)) = \begin{cases} -x_{\max}, & x(t) < -x_{\max} \\ x(t), & |x(t)| \le x_{\max} \\ x_{\max}, & x(t) > x_{\max} \end{cases}, \quad (24)$$

where $x_{\max}$ is the clipping threshold. The soft saturation model is expressed as

$$f(x(t)) = \frac{x_{\max}x(t)}{\sqrt[\rho]{|x_{\max}|^\rho + |x(t)|^\rho}}, \quad (25)$$

where $\rho$ is a nonadaptive shape parameter.

For the implementation of the algorithms, both matched and unmatched conditions are considered. In the matched condition, the same nonlinearity is used as both the actual model and the numerical model in the algorithms. In the unmatched condition, the nonlinear expansion in (2) is used as the numerical model. In all the simulations and experiments, the nonlinear expansion order of the algorithms is set as $p = 3$.

*1) Single-Talk Case*

A 10-s long female speech signal as the far-end input signal $x(t)$ is distorted using the two nonlinear mappings to generate the microphone signal $y(t)$ following (1), and white Gaussian noise $s(t)$ is utilized to represent the background noise in the single-talk case. We use the signal-to-distortion ratio (SDR) to quantify the degree of nonlinearity, which is defined as $10\log_{10}\{E[x^2(t)]/E[(f(x(t))-x(t))^2]\}$ [23], and it is set to 5 dB. The echo-to-near-end-signal power ratio (ESR), defined as $10\log_{10}\{E[d^2(t)]/E[s^2(t)]\}$ [23], is set as ESR = 60 dB. The echo path is a room impulse response generated by the image method [32] with a sampling rate of 16 kHz and a reverberation time of 0.2 s. The learning rate $\eta$ of the SBSS algorithm is set to 0.1. The performance is measured by the echo return loss enhancement (ERLE), defined as $10\log_{10}\{E[y^2(t)]/E[e^2(t)]\}$ [25]. The ERLE results in the two nonlinear cases are shown in Fig. 2. It can be seen that the SD-MCSSFDAF and the DualStage-MCK algorithms significantly outperform the SBSS algorithm after convergence when the nonlinearity is perfectly matched. However, when the nonlinear expansion is utilized, these two algorithms deteriorate considerably due to the influence of the mismatch of nonlinearity, and the benefit of the proposed nonlinear SBSS method can be clearly seen.

*2) Double-Talk Case*

The far-end speech signal $x(t)$ is again distorted using the two nonlinear mappings with SDR = 5 dB. The near-end signal $s(t)$, which is a 10-s long male speech signal, is superimposed on the echo signal $d(t)$ to obtain the microphone signal $y(t)$ with ESR = 0 dB. Only more practical unmatched condition is considered in the double-talk case, and the performance is measured by the true ERLE (tERLE), defined as $10\log_{10}\{E[d^2(t)]/E[(e(t)-s(t))^2]\}$ [7]. Besides, perceptual evaluation of speech quality (PESQ) [33] and short-time objective intelligibility (STOI) [34], [35] are also employed as objective measures to evaluate the speech quality of the near-end signal. The tERLE results of the two nonlinear mappings are shown in Fig. 3, and the evaluation results of the near-end speech quality with different SDR values are shown in Table I, from which the efficacy of the proposed method can be seen.

### B. Real Experiments

We also evaluate the performance of the proposed SBSS algorithm using real captured data. A male speech signal emitted by a miniature loudspeaker, which inevitably includes an unknown nonlinearity, is recorded by a microphone with signal-to-noise ratio (SNR) of approximately 20 dB. The length of the signal is 10 s long, with a sampling rate of 16 kHz. Fig. 4 shows the ERLE results for the single-talk case. Obviously, the ERLE performance of the SBSS algorithm is better than that of the other two algorithms in this unmatched condition. In the case of double-talk, a 10-s long female speech signal is used as the near-end signal, and the volume is adjusted to achieve an ESR of 0 dB. The tERLE results are shown in Fig. 5, and the evaluation results of the near-end speech quality with different SNR values are shown in Table II. It can be seen that the SBSS algorithm not only achieves the best echo cancellation performance in double-talk condition, but also has the best near-end speech quality.

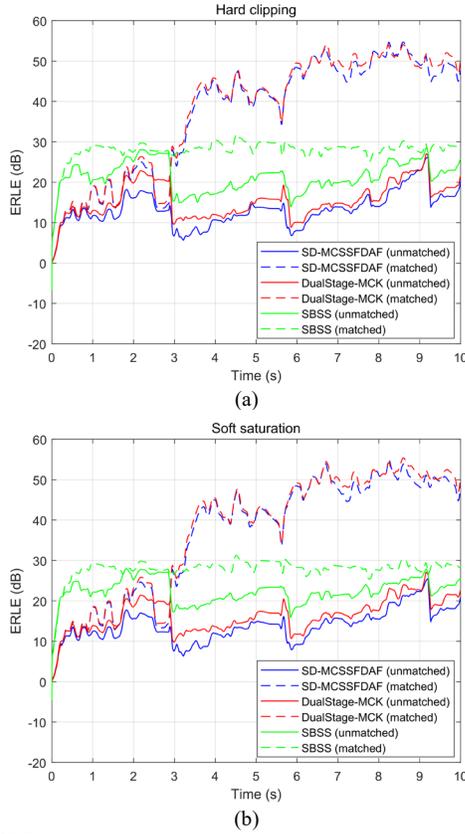

Fig. 2. ERLE results in matched and unmatched conditions. (a) Hard clipping. (b) Soft saturation.

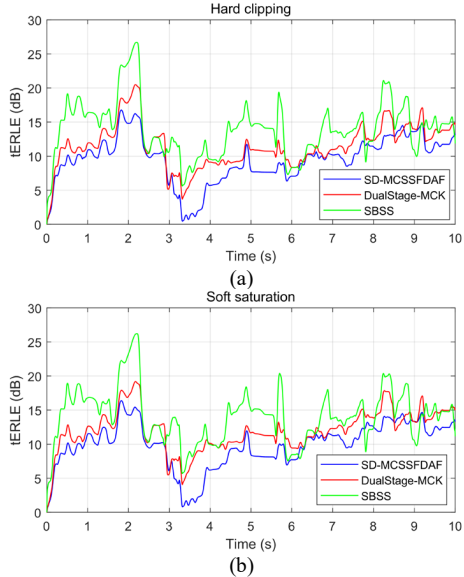

Fig. 3. tERLE results in unmatched condition. (a) Hard clipping. (b) Soft saturation.

TABLE I
EVALUATION RESULTS OF NEAR-END SPEECH QUALITY

| SDR | Algorithms | Hard clipping | | Soft saturation | |
|---|---|---|---|---|---|
| | | PESQ | STOI | PESQ | STOI |
| 3 dB | SD-MCSSFDAF | 1.24 | 0.80 | 1.27 | 0.82 |
| | DualStage-MCK | 1.29 | 0.82 | 1.33 | 0.84 |
| | SBSS | 1.38 | 0.85 | 1.43 | 0.87 |
| 5 dB | SD-MCSSFDAF | 1.43 | 0.86 | 1.47 | 0.87 |
| | DualStage-MCK | 1.53 | 0.88 | 1.59 | 0.89 |
| | SBSS | 1.69 | 0.91 | 1.80 | 0.92 |
| 10 dB | SD-MCSSFDAF | 1.82 | 0.93 | 1.84 | 0.93 |
| | DualStage-MCK | 2.06 | 0.95 | 2.13 | 0.96 |
| | SBSS | 2.35 | 0.96 | 2.51 | 0.97 |

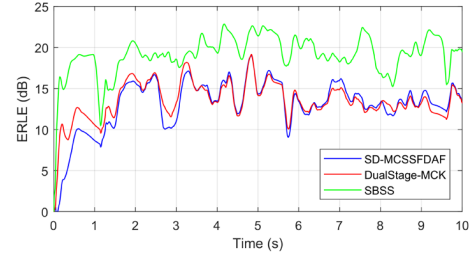

Fig. 4. ERLE results of real data for the single-talk case.

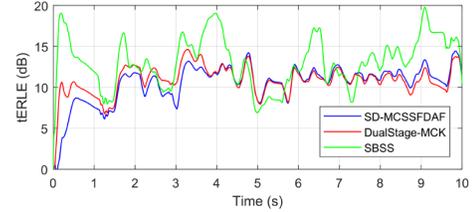

Fig. 5. tERLE results of real data for the double-talk case.

TABLE II
EVALUATION RESULTS OF NEAR-END SPEECH QUALITY USING REAL DATA

| Algorithms | SNR = 10 dB | | SNR = 15 dB | | SNR = 20 dB | |
|---|---|---|---|---|---|---|
| | PESQ | STOI | PESQ | STOI | PESQ | STOI |
| SD-MCSSFDAF | 1.42 | 0.80 | 1.51 | 0.82 | 1.56 | 0.83 |
| DualStage-MCK | 1.45 | 0.81 | 1.53 | 0.83 | 1.58 | 0.84 |
| SBSS | 1.52 | 0.83 | 1.61 | 0.85 | 1.67 | 0.86 |

## V. CONCLUSION

In this letter, we propose a novel NAEC algorithm based on SBSS. We merge the nonlinear expansion coefficients of the basis functions into the echo path. By regarding all the basis functions of the far-end input signal as the known equivalent reference signals, an online SBSS algorithm can be derived using the constrained scaled natural gradient strategy. The proposed SBSS algorithm, based on the independence between the near-end signal and the reference signals, is less sensitive to the mismatch of nonlinearity between the numerical and actual models than the NAEC algorithm based on adaptive filtering. Simulations using two types of nonlinear mappings and experiments using real captured data validate that the proposed SBSS algorithm achieves effective nonlinear echo cancellation performance when the numerical nonlinear model mismatches the actual model.

## References

[1] E. Hänsler and G. Schmidt, *Acoustic Echo and Noise Control: A Practical Approach*. Hoboken, NJ, USA: Wiley, 2004.

[2] H. Zhao, Y. Yu, S. Gao, X. Zeng and Z. He, "Memory proportionate APA with individual activation factors for acoustic echo cancellation," *IEEE/ACM Trans. Audio, Speech, Lang. Process.*, vol. 22, no. 6, pp. 1047–1055, Jun. 2014.

[3] W. Fan, K. Chen, J. Lu and J. Tao, "Effective improvement of under-modeling frequency-domain Kalman filter," *IEEE Signal Process. Lett.*, vol. 26, no. 2, pp. 342–346, Feb. 2019.

[4] J. F. Cardoso, "Blind signal separation: statistical principles," *Proceedings of the IEEE*, vol. 86, no. 10, pp. 2009–2025, Oct. 1998.

[5] S. Miyabe, T. Takatani, H. Saruwatari, K. Shikano, and Y. Tatekura, "Barge-in and noise-free spoken dialogue interface based on sound field control and semi-blind source separation," in *Proc. Eur. Signal Process. Conf.*, Florence, Italy, Sep. 2007, pp. 232–236.

[6] T. S. Wada, S. Miyabe, and B. H. Juang, "Use of decorrelation procedure for source and echo suppression," in *Proc. IWAENC*, Seattle, WA, Sep. 2008.

[7] F. Nesta, T. S. Wada, and B. H. Juang, "Batch-online semi-blind source separation applied to multi-channel acoustic echo cancellation," *IEEE Trans. Audio, Speech, Lang. Process.*, vol. 19, no. 3, pp. 583–599, Mar. 2011.

[8] M. Joho, H. Mathis, and G. S. Moschytz, "Combined blind/nonblind source separation based on the natural gradient," *IEEE Signal Process. Lett.*, vol. 8, no. 8, pp. 236–238, Aug. 2001.

[9] J. Gunther, "Learning echo paths during continuous double-talk using semi-blind source separation," *IEEE Trans. Audio, Speech, Lang. Process.*, vol. 20, no. 2, pp. 646–660, Feb. 2012.

[10] Z. Koldovský, J. Málek, M. Müller, and P. Tichavský, "On semi-blind estimation of echo paths during double-talk based on nonstationarity," in *Proc. IWAENC*, Juan-les-Pins, France, 2014, pp. 198–202.

[11] J. Gunther and T. Moon, "Blind acoustic echo cancellation without double-talk detection," in *Proc. IEEE Workshop Appl. Signal Process. Audio Acoust.*, New Paltz, NY, USA, Oct. 2015, pp. 1–5.

[12] M. Zeller and W. Kellermann, "Coefficient pruning for higher-order diagonals of Volterra filters representing Wiener-Hammerstein models," in *Proc. Int. Workshop, Acoust. Echo, Noise Control*, Seattle, WA, Sep. 2008.

[13] M. Zeller and W. Kellermann, "Fast and robust adaptation of DFT-domain Volterra filters in diagonal coordinates using iterated coefficient updates," *IEEE Trans. Signal Process.*, vol. 58, no. 3, pp. 1589–1604, Mar. 2010.

[14] B. S. Nollett and D. L. Jones, "Nonlinear echo cancellation for hands-free speakerphones," in *Proc. IEEE Workshop, Nonlinear Signal, Image Process.*, Mackinac Island, MI, Sep. 1997.

[15] A. N. Birkett and R. A. Goubran, "Acoustic echo cancellation using NLMS-neural network structures," in *Proc. IEEE Int. Conf. Acoust., Speech, Signal Process.*, Detroit, MI, May 1995, vol. 5, pp. 3035–3038.

[16] F. Küch, A. Mitnacht, and W. Kellermann, "Nonlinear acoustic echo cancellation using adaptive orthogonalized power filters," in *Proc. IEEE Int. Conf. Acoust., Speech, Signal Process.*, Philadelphia, PA, Mar. 2005, vol. 3, pp. 105–108.

[17] A. Stenger and W. Kellermann, "Adaptation of a memoryless preprocessor for nonlinear acoustic echo cancelling," *Signal Process.*, vol. 80, no. 9, pp. 1747–1760, Sep. 2000.

[18] J. P. Costa, A. Lagrange, and A. Arliaud, "Acoustic echo cancellation using nonlinear cascade filters," in *Proc. IEEE Int. Conf. Acoust., Speech, Signal Process.*, Hong Kong, China, Apr. 2003, vol. 5, pp. 389–392.

[19] D. Comminiello, M. Scarpiniti, L. A. Azpicueta-Ruiz, J. Arenas-García, and A. Uncini, "Functional link adaptive filters for nonlinear acoustic echo cancellation," *IEEE Trans. Audio, Speech, Lang. Process.*, vol. 21, no. 7, pp. 1502–1512, Jul. 2013.

[20] C. Hofmann, C. Huemmer, and W. Kellermann, "Significance-aware Hammerstein group models for nonlinear acoustic echo cancellation," in *Proc. IEEE Int. Conf. Acoust., Speech, Signal Process.*, Florence, Italy, May 2014, pp. 5934–5938.

[21] C. Hofmann, M. Guenther, C. Huemmer, and W. Kellermann, "Efficient nonlinear acoustic echo cancellation by partitioned-block significance-aware Hammerstein group models," in *Proc. Eur. Signal Process. Conf.*, Budapest, Hungary, Aug. 2016, pp. 1783–1787.

[22] S. Malik and G. Enzner, "Fourier expansion of Hammerstein models for nonlinear acoustic system identification," in *Proc. IEEE Int. Conf. Acoust., Speech, Signal Process.*, Prague, CZ, May 2011, pp. 85–88.

[23] S. Malik and G. Enzner, "State-space frequency-domain adaptive filtering for nonlinear acoustic echo cancellation," *IEEE Trans. Audio, Speech, Lang. Process.*, vol. 20, no. 7, pp. 2065–2079, Sep. 2012.

[24] G. Enzner and P. Vary, "Frequency-domain adaptive Kalman filter for acoustic echo control in hands-free telephones," *Signal Process.*, vol. 86, no. 6, pp. 1140–1156, Jun. 2006.

[25] J. Park and J. Chang, "State-space microphone array nonlinear acoustic echo cancellation using multi-microphone near-end speech covariance," *IEEE/ACM Trans. Audio, Speech, Lang. Process.*, vol. 27, no. 10, pp. 1520–1534, Oct. 2019.

[26] M. Schrammen, S. Kühl, S. Markovich-Golan, and P. Jax, "Efficient nonlinear acoustic echo cancellation by dual-stage multi-channel kalman filtering," in *Proc. IEEE Int. Conf. Acoust., Speech Signal Process.*, Brighton, United Kingdom, May 2019, pp. 975–979.

[27] S. Douglas and M. Gupta, "Scaled natural gradient algorithms for instantaneous and convolutive blind source separation," in *Proc. ICASSP*, Apr. 2007, vol. II, pp. 637–640.

[28] T. Kim, H. T. Attias, S.-Y. Lee, and T.-W. Lee, "Blind source separation exploiting higher-order frequency dependencies," *IEEE Trans. Audio, Speech, Lang. Process.*, vol. 15, no. 1, pp. 70–79, Jan. 2007.

[29] T. Kim, "Real-time independent vector analysis for convolutive blind source separation," *IEEE Trans. on Circuit and systems*, vol. 57, no. 7, pp. 1431–1438, Jul. 2010.

[30] S. Erateb, M. Naqvi, and J. Chambers, "Online IVA with adaptive learning for speech separation using various source priors," in *SSPD*, London, UK, 2017, pp. 74–78.

[31] S. Erateb and J. Chambers, "Enhanced online IVA with switched source prior for speech separation," in *IEEE SAM*, Hangzhou, China, 2020, pp. 1–5.

[32] J. B. Allen and D. A. Berkley, "Image method for efficiently simulating small-room acoustics," *J. Acoust. Soc. Amer.*, vol. 65, no. 4, pp. 943–950, Apr. 1979.

[33] ITU-T, *Perceptual evaluation of speech quality (PESQ): An objective method for end-to-end speech quality assessment of narrow-band telephone networks and speech codecs*, Rec. 862, International Telecommunications Union, 2000.

[34] C. H. Taal, R. C. Hendriks, R. Heusdens, and J. Jensen, "A short-time objective intelligibility measure for time-frequency weighted noisy speech," in *ICASSP*, Dallas, TX, USA, Mar. 2010, pp. 4214–4217.

[35] C. H. Taal, R. C. Hendriks, R. Heusdens, and J. Jensen, "An algorithm for intelligibility prediction of time–frequency weighted noisy speech," *IEEE Trans. Audio, Speech, Lang. Process.*, vol. 19, no. 7, pp. 2125–2136, Sep. 2011.